\documentclass[preprint,12pt,a4paper]{elsarticle}
\usepackage{amssymb}
\usepackage{array}
\newcolumntype{L}[1]{>{\raggedright\arraybackslash}p{#1}}
\usepackage{graphicx}
\usepackage{float}
\usepackage{hyperref}
\journal{SoftwareX}

\begin{document}
\begin{frontmatter}

\title{EREBOTS 4.0: A framework for conversational assistants with A2A agents and MCP tools}
\author[hevs,uex]{Elia Pacioni\corref{cor1}}
\ead{elia.pacioni@hevs.ch}

\author[hevs]{Davide Calvaresi}
\author[hevs]{Andrei C. Coman}
\author[hevs]{Ismael Jaggi}
\author[hevs]{Valerio Crocetti}
\author[hevs]{Gaetano Manzo}
\author[hevs]{Michael Schumacher}

\address[hevs]{HES-SO Valais-Wallis, Rue de Technopôle 10, 3960 Sierre, Switzerland}
\address[uex]{University of Extremadura, 06800, M\'erida, Spain}

\cortext[cor1]{Corresponding author.}

\begin{abstract}
In recent years, multi-agent systems have shifted from rule-based, symbolic agents to sub-symbolic agents built on language models. However, several limitations remain unresolved. The new frontier is neuro-symbolic agents, which pair language models with deterministic tools and explicit protocols. Without a common framework to standardize and compose them, each study rebuilds the same infrastructure, and results are hard to replicate. This paper presents EREBOTS 4.0, an open-source, hybrid-AI, multi-agent framework that fills this gap for conversational assistants. Language-model agents plan and route requests, while deterministic tools compute exact results over the Model Context Protocol (MCP). An orchestrator selects one or more specialist agents per turn and calls them over the Agent2Agent protocol (A2A). A new agent, local or remote, is added by listing its URL in a configuration file. The framework also manages identity, user context with consent flags, task states and traces. Its A2A server passed 97 tracked requirements in the official A2A Technology Compatibility Kit, with no failures.
\end{abstract}

\end{frontmatter}

\section*{Required Metadata}

\section*{Current code version}

Table~\ref{tab:metadata} describes the version reported here.

\begin{table}[!ht]
\small
\begin{tabular}{|l|p{5.8cm}|p{5.8cm}|}
\hline
\textbf{Nr.} & \textbf{Code metadata description} & \textbf{Value} \\
\hline
C1 & Current code version & 1.0.0 \\
\hline
C2 & Permanent GitHub link to code/repository used for this code version & \url{https://github.com/aislab-hevs/erebots4.0/tree/v1.0.0} \\
\hline
C3 & Legal Code License & MIT \\
\hline
C4 & Code versioning system used & git \\
\hline
C5 & Software code languages, tools, and services used & Python, FastAPI, \texttt{a2a-sdk}, MCP Python SDK, OpenAI-compatible LLM APIs, SQLAlchemy, PostgreSQL and SQLite, Keycloak, Vue.js, Gradio, Docker Compose, Traefik \\
\hline
C6 & Compilation requirements, operating environments \& dependencies & Python $\geq$ 3.13 and an OpenAI-compatible LLM endpoint. A checkout resolves from a \texttt{uv} lockfile, and the published wheel bounds the three SDKs the framework codes against. Tool-using specialists need a model that supports tool calls. Docker Compose provides an optional deployment stack \\
\hline
C7 & If available Link to developer documentation/manual & \url{https://github.com/aislab-hevs/erebots4.0/tree/v1.0.0/docs} \\
\hline
C8 & Support email for questions & \href{mailto:elia.pacioni@hevs.ch}{elia.pacioni@hevs.ch} \\
\hline
\end{tabular}
\caption{Code metadata}
\label{tab:metadata}
\end{table}

\section{Motivation and significance}
\label{sec:motivation}

The building blocks of a multi-agent system have changed. Most multi-agent research built its agents from deliberative components that exchanged explicit communication acts, coordinating or competing over goals that were their own or partly shared~\cite{wooldridge1995}. A current wave builds them from language models that plan, call tools, and delegate in natural language, and the surveys of that shift already run to hundreds of primary studies~\cite{calvaresi2023slr,pacioni2026slr}. One reading of this wave is that it renames the earlier concepts~\cite{botti2025}. The engineering problem moved with that shift, from building an agent to composing several. An agent is now assembled from parts that each hold a model, a prompt, and a set of executable operations, and those parts are often written by different groups on different hosts.

Research on such systems needs a platform, for the same reason earlier multi-agent research needed one. Comparing routing strategies, delegation policies, or the right granularity for a specialist means holding everything else fixed: discovery, transport, task lifecycle, streaming, identity, conversation state. A researcher who rebuilds that substrate for each study spends the effort there, and the result is hard for anyone else to replicate. A platform that holds it fixed turns those questions into things a researcher can change one at a time and measure, and into claims others can test on the same substrate.

The challenge itself is not new. JADE~\cite{bellifemine2007jade} and SPADE~\cite{palanca2020spade} gave multi-agent research platforms where agents met over a specification, FIPA-ACL~\cite{fipa2002acl} in one case and XMPP~\cite{rfc6120} in the other, so researchers could compose each other's agents and replicate an experiment elsewhere. Rasa played a comparable role for open conversational systems~\cite{bocklisch2017rasa}, and PEAK still builds agent communities on SPADE today~\cite{ribeiro2025peak}. 

Two specifications now do that job for language-model agents. A2A defines agent cards, messages, artifacts, and task operations~\cite{a2a2025}, and MCP defines tool discovery and invocation~\cite{mcp2026}. Both have been taken up by the frameworks in common use.

LangGraph~\cite{langgraph}, CrewAI~\cite{crewai}, AutoGen~\cite{wu2023autogen}, Microsoft Agent Framework~\cite{microsoft2025agentframework} and Google ADK~\cite{adk} are toolkits for building and orchestrating agents. They provide stores for conversation history and memory, and leave it to the application to authenticate the user and tie each request to that user. All are permissively licensed, four under MIT terms and ADK under Apache 2.0. Several of them now speak A2A. CrewAI delegates to a remote agent from its card URL, and Google ADK and Microsoft Agent Framework wrap one the same way. A LangGraph assistant serves a card of its own once deployed on LangChain's hosted server~\cite{langchain}. In all of them the peer is reached from application code, so adding a peer or moving it to another host means changing that code. CrewAI and LangChain develop their open core alongside a commercial managed platform. AutoGen began as a research artifact, and its repository now describes it as in maintenance mode, with no new features and under community management.

EREBOTS 4.0 is a new generation of the EREBOTS platform for personalized health chatbots~\cite{calvaresi2021erebots}, which supported work on nutritional coaching~\cite{calvaresi2021nutrition} and explainable assistants~\cite{buzcu2024framework}. This implementation introduces LLM-driven routing and specialists behind protocol endpoints. EREBOTS 4.0 is built for one kind of system, an assistant a person talks to. That narrowing is what lets the framework take over what those toolkits leave to the application, since every assistant needs it. The orchestrator takes the identity from the login, loads the context stored for that user, and passes both to every specialist it calls, local or remote, in A2A metadata. The examples use personal-data management and a deterministic calculation, which explain delegation, tool execution, and conversational state. It declares its routable specialists as A2A endpoints (URL) in a configuration file, and the orchestrator discovers what each offers from its agent card without application code constructing a wrapper for it. Writing an agent here means subclassing a single class and naming the MCP servers that hold its tools.

A researcher uses the framework by writing a configuration file that lists agent endpoints, registering any specialist hosted in their own application, pointing each agent at a model endpoint, and starting the application with a GUI or CLI.

This paper presents three main contributions (C):

\begin{itemize}
\item C1. a framework that composes specialists over A2A and MCP, so they run in one process or on separate hosts with no code change;
\item C2. the application layer a deployed study needs beyond an agent library, with OpenID Connect (OIDC) login, user-context persistence, execution traces, and one conversation service every frontend shares;
\item C3. a measured conformance result against the official A2A Technology Compatibility Kit, reported with the requirements it leaves untested and the two \texttt{a2a-sdk} deviations the measurement exposed.
\end{itemize}

\section{Software description}
\label{sec:description}

The repository separates the installable \texttt{erebots} package from application composition, specialist implementations, and frontend examples. Application code constructs agents, configures persistence and credentials, and mounts their endpoints. The core can also serve an orchestrator with no registered specialists. The examples in this paper use two specialists: \texttt{PersonalAgent} manages a user's own data, and \texttt{ConversionAgent} converts between units.

\subsection{Software architecture}
\label{sec:architecture}

\begin{figure}[!ht]
\centering
\includegraphics[width=\linewidth]{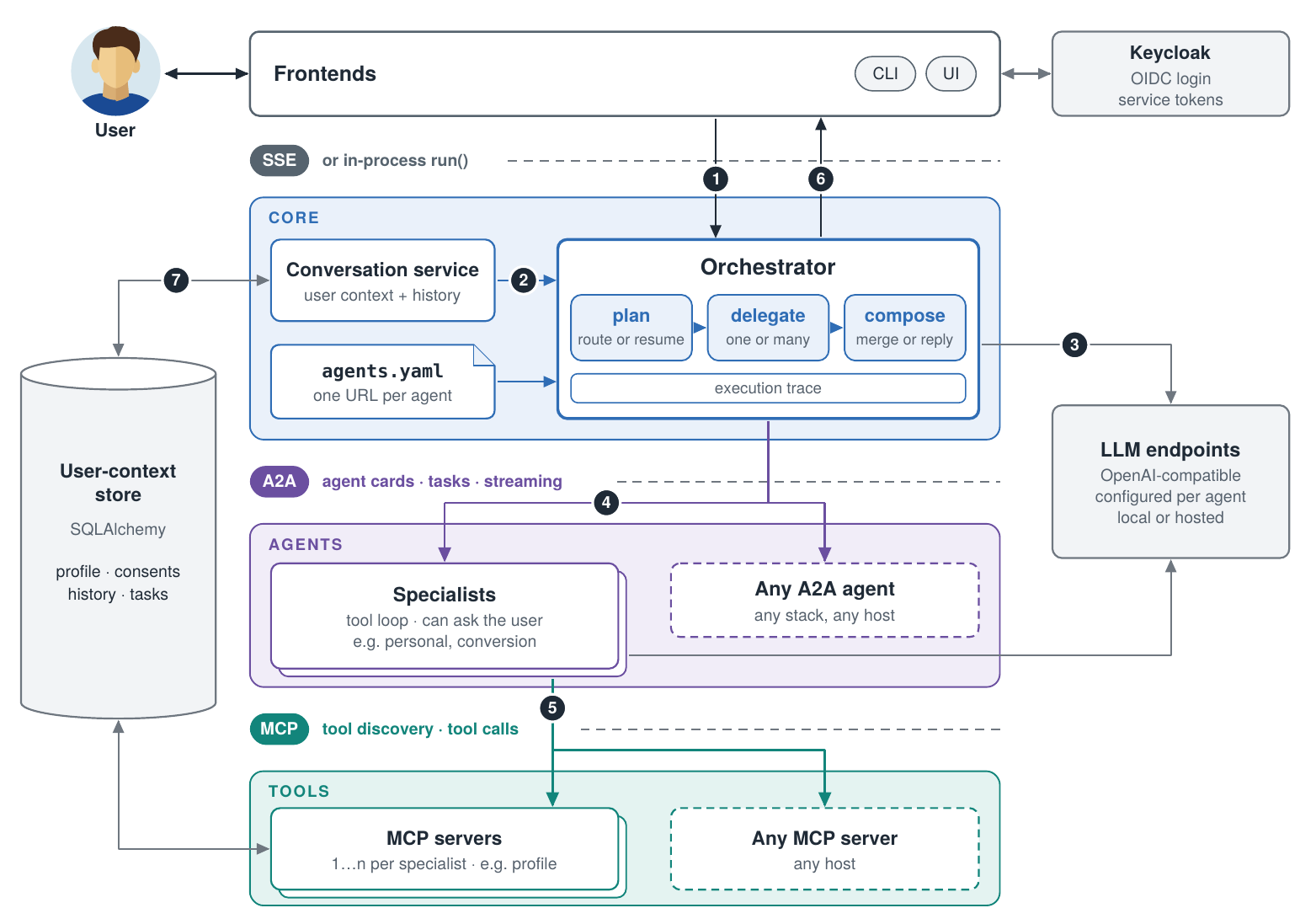}
\caption{EREBOTS 4.0 architecture. One user turn: (1) message, (2) context loading, (3) LLM planning, (4) A2A delegation, (5) MCP tool calls, (6) streamed answer, (7) persistence. Dashed lines: possible process or host boundaries.}
\label{fig:architecture}
\end{figure}

This section follows one interaction, from the user's request through the tools to the answer. Figure~\ref{fig:architecture} shows the architecture. A frontend supplies the message and the session context. The orchestrator streams the turn back as it runs, carrying the routing decision, answer and reasoning tokens, trace steps, and the outcome, which is an answer, an error, or an \texttt{input\allowbreak\_required} request for a missing value. A separate channel carries a contributing specialist's output, so the user reads a single answer. A frontend in the same process consumes the stream directly, and one across the network receives it as Server-Sent Events (SSE).

The application lists routable endpoints in \texttt{agents.yaml}. The path can be passed in, and otherwise resolves from the working directory, without requiring a source checkout. The orchestrator then fetches each agent card and reads the name, description, skills, and endpoint it advertises. A routing prompt presents these capabilities to the orchestrator's LLM, together with recent conversation turns for follow-up messages. The selected names are checked against the registry and bounded by a configured maximum. An empty selection produces a direct response from the orchestrator. What the orchestrator does next depends on how many specialists it selected. For one selected specialist, the orchestrator relays its streamed answer. For several, it starts concurrent A2A calls and tags each contribution with its agent name on a separate channel. Each specialist receives the request and a description of the responsibilities assigned to the others. The orchestrator then synthesizes the contributions with an additional LLM call, or concatenates them into sections.

Concurrent calls change what a failure costs. A failing call removes one contribution and leaves the rest available. The trace records it and the answer omits it, unless every call fails and the turn ends in an error naming each. Concurrent specialists all receive the context prepared at the start of the turn, so an update made by one is not visible to calls already in progress. Specialists do not call one another. The orchestrator holds the only A2A client, and the interaction graph is a star.

Specialists written inside the framework inherit \texttt{BaseSpecialist}, which implements MCP discovery and a bounded cycle of model calls, tool invocations, and tool results. Its A2A skills say what it can be asked for, and its MCP tool schemas say what it can execute. A tool that runs and reports failure is distinct from a server that cannot be reached, and the trace records each as its own step.

A specialist does not always have what it needs to answer. The built-in \texttt{ask\_user} control ends a streaming iteration with a request for missing information, and the A2A executor marks the task \texttt{input-required}. A client resumes by sending the answer and the task identifier. The server reconstructs the exchange from the message that opened the task, so the specialist sees the question it asked and the context that preceded it. The orchestrator then bypasses routing, after checking ownership against the requesting identity and checking expiry. The mechanism preserves task identity and does not checkpoint the internal LLM tool loop. A resumed specialist no longer knows which tools it already called, so any tool it might reach before asking the question has to be safe to run twice.

A turn that crosses the orchestrator, a specialist, and a tool server cannot be reconstructed from any one of them. Traces record routing decisions, delegations, tool calls, tool results, and tool failures. The failure taxonomy for multi-agent LLM systems was built from more than 1600 such traces, annotated across seven frameworks~\cite{cemri2025}. Step identifiers and parent identifiers associate a specialist's trace with the A2A call that produced it. Durations and token counts are attached where collected. These records describe observable execution events and do not establish the causal reasoning behind a model's choice. Context propagation and trace payloads travel inside A2A messages and artifacts as EREBOTS 4.0 metadata conventions.

\subsection{Software functionalities}
\label{sec:extension}

The A2A server layer uses the official \texttt{a2a-sdk} and exposes JSON-RPC and HTTP+JSON routes from one request handler, with streaming, task operations, and push notifications. An extended agent card is advertised when one is configured. Push delivery validates the webhook address at registration and again at delivery, then pins the connection to the validated address. The default task store is in memory, and the application can select database-backed task and push-configuration stores. A parked task then survives a restart, verified by recreating the server, the executor, and the store and resuming from the task identifier alone.

An identity enters through Keycloak~\cite{keycloak}, which provides OIDC login over cookies or bearer tokens. The application derives it from the authenticated session and hands it to the specialist. There \texttt{BaseSpecialist} exposes a \texttt{prepare\_tool\_args} hook for pinning server-side arguments, and the \texttt{PersonalAgent} uses it to overwrite the \texttt{user\_id} of every tool call before the call runs, so a model cannot address another user's record.

Internal traffic carries its own credentials. A2A and MCP routes accept service-account tokens obtained through \texttt{client\_credentials}, and check which client each token was issued to. Every request below \texttt{/tools/} carries that bearer, including reads from the loopback interface. An application can opt into one exception, which exists for a single UI library and defaults to off. Two gaps remain in that chain. Internal services are trusted to propagate the identity correctly, and the command-line client selects one with \texttt{--user} without authenticating it, which makes it a development tool.

The \texttt{PersonalAgent} exposes profile updates, consent updates, and context retrieval through MCP, over a SQLAlchemy store. It keeps consent flags there and serves them to the application, which decides what to do with them. The same store holds the transcript, reached through a conversation-service interface that every frontend shares. That interface owns the transcript when an identity and a store are both present, and reports that it does not when either is missing. Appending a turn is a single transaction whose first statement is a write, which serializes two clients answering at once and loses neither turn. The suite exercises that path on SQLite.

Local agents are registered in \texttt{specialists/}\allowbreak\texttt{registry.py}. A specialist supplies a prompt, a description, skills, and the addresses of the MCP servers holding its tools. The registry returns it under a name, and bootstrap mounts its A2A application at the matching path. That path is what \texttt{agents.yaml} lists:

\begin{verbatim}
agents:
  - http://localhost:7860/a2a/personal
\end{verbatim}

The example application mounts every specialist and every tool server in one Asynchronous Server Gateway Interface (ASGI) application, so one port serves the whole assistant on a laptop, while the conformance fixtures run in their own process with their own dependency set.

An independently hosted specialist supplies its own implementation and endpoint, and the orchestrator needs no domain import for it. Connectivity, authentication, supported message forms, and any shared metadata conventions must still match. Model endpoints are configured per agent, tool-using agents require tool-call support, and image inputs require a multimodal deployment. For in-process testing, the orchestrator accepts an injected A2A client, so a consumer can point it at an ASGI application without opening a socket.

\subsection{Verification}
\label{sec:verification}

We measured EREBOTS4.0, release 1.0.0, with an automated suite and a conformance run, under Python 3.13~\cite{python313} with \texttt{a2a-sdk} 1.1.2, \texttt{mcp} 1.29.1, and \texttt{openai} 2.54.0. The suite passes 567 tests covering routing and fan-out with stubbed model calls, A2A round-trips through the SDK, task parking and resume across a recreated store, MCP discovery and tool-error handling, authentication, persistence under concurrent writes, and frontend event handling.

The A2A Technology Compatibility Kit (TCK) grades a server against the capabilities its own agent card declares. Run against the fixture endpoints of version 1.0.0, built on \texttt{a2a-sdk} 1.1.2 and serving A2A v1.0, it reports no failures at MUST, SHOULD, or MAY level~\cite{rfc2119}. The fixtures answer from canned payloads with no model in the loop. One endpoint declares every optional capability. \texttt{/a2a/tck-minimal} serves the same scenarios behind a card that declares those capabilities absent. Only there can the suite reach \texttt{CORE-CAP-001} to \texttt{003}, which specify the error a server must return when a caller asks for a capability it does not declare.

Table~\ref{tab:tck} gives the counts. The 94 requirements passed on \texttt{/a2a/tck} are 83 MUST, 7 SHOULD, and 4 MAY, and the union of 97 adds the three \texttt{CORE-CAP} requirements only the minimal card reaches. Both runs omitted the transport flag, and each covered the JSON-RPC and HTTP+JSON bindings the cards declare. Table~\ref{tab:tck} aggregates across bindings, and the published reports hold the per-binding split.

\begin{table}[!ht]
\centering
\small
\begin{tabular}{|L{0.40\linewidth}|r|r|r|r|}
\hline
Evaluation unit & Passed & Failed & Skipped & Not tested \\
\hline
Tests, \texttt{/a2a/tck} & 180 & 0 & 85 & n/a \\
\hline
Tests, \texttt{/a2a/tck-minimal} & 140 & 0 & 125 & n/a \\
\hline
Requirements, \texttt{/a2a/tck} & 94 & 0 & 10 & 25 \\
\hline
Requirements, \texttt{/a2a/tck-minimal} & 73 & 0 & 30 & 26 \\
\hline
Requirements, union of both runs & 97 & 0 & 7 & 25 \\
\hline
\end{tabular}
\caption{TCK counts for version 1.0.0, out of the 129 requirements the suite tracks. The two endpoints cover different requirements by construction. The minimal card declares three capabilities absent, which is why 30 requirements are skipped against it, and without streaming it offers no Server-Sent Events to exercise.}
\label{tab:tck}
\end{table}

Seven requirements are skipped across both runs. Six concern the gRPC binding, which the server neither serves nor declares, and the seventh applies to a server that declares an extended card without configuring it, a state the card builder cannot produce. A further 25 have no test in the suite at all. Zero failures means that every exercised requirement passed, not that every normative requirement was tested.

Reaching zero failures required correcting two deviations from MUST requirements in \texttt{a2a-sdk} 1.1.2. On the HTTP+JSON binding it answers \texttt{Task\allowbreak NotCancelable\allowbreak Error} with 400 where the specification's error table gives 409 (\texttt{CORE-CANCEL-002}). The JSON-RPC binding is correct because it reads a separate table. It also answers a subscription to a terminal task with \texttt{Invalid\allowbreak Params\allowbreak Error} where the specification requires \texttt{Unsupported\allowbreak Operation\allowbreak Error} (\texttt{STREAM-SUB-003}). The second is a regression between the SDK's v1 and v2 handlers. The server layer overrides both, so the runs reported here pass those requirements, and any deployment built on that SDK version inherits the defects without the overrides.

We checked the distribution the way a consumer meets it, installing the built wheel outside the repository and implementing an agent from the documentation alone. That found a defect the in-repository suite could not reach. \texttt{MCPClient.\allowbreak call\_tool} ignored the \texttt{isError} flag of an MCP response and returned the server's error text as a result. The version reported here raises \texttt{MCPToolError} instead.

\section{Illustrative examples}
\label{sec:examples}

We ship a demonstrator that starts \texttt{PersonalAgent} together with a \texttt{ConversionAgent} whose single MCP tool converts quantities between units of mass, length, energy, and volume by exact factors. The following two scenarios distinguish framework mechanics from model decisions. Routing dispatch, the A2A call, the tool invocation, the stored value, and the task lifecycle are pinned by the automated suite with stubbed model replies, so they do not depend on which model is deployed. The choice of specialist and the decision to ask the user depend on the model.

A user writes \emph{``Hello, my preferred language is Italian.''} The intended route is \texttt{PersonalAgent}, which calls the profile-update tool, and reading the value back with the context-retrieval operation is what establishes success, since a generated confirmation does not establish that the profile changed. The user then writes \emph{``I made a mistake. Update my preferred language.''} The specialist calls \texttt{ask\_user}, the task becomes \texttt{input-required}, and the user answers \emph{``My preferred language is Spanish.''} The reply carries the task identifier and no transcript, and the profile changes on that second turn. No system prompt names \texttt{ask\_user}, and the framework exposes no \texttt{tool\_choice} setting, so the tool description is the only instruction the model receives about it. A parked task resumes after the server and its store are recreated, the same resume is refused to a different identity, and it is refused once the task has expired. 

\begin{figure}[H]
\centering
\includegraphics[width=0.95\linewidth]{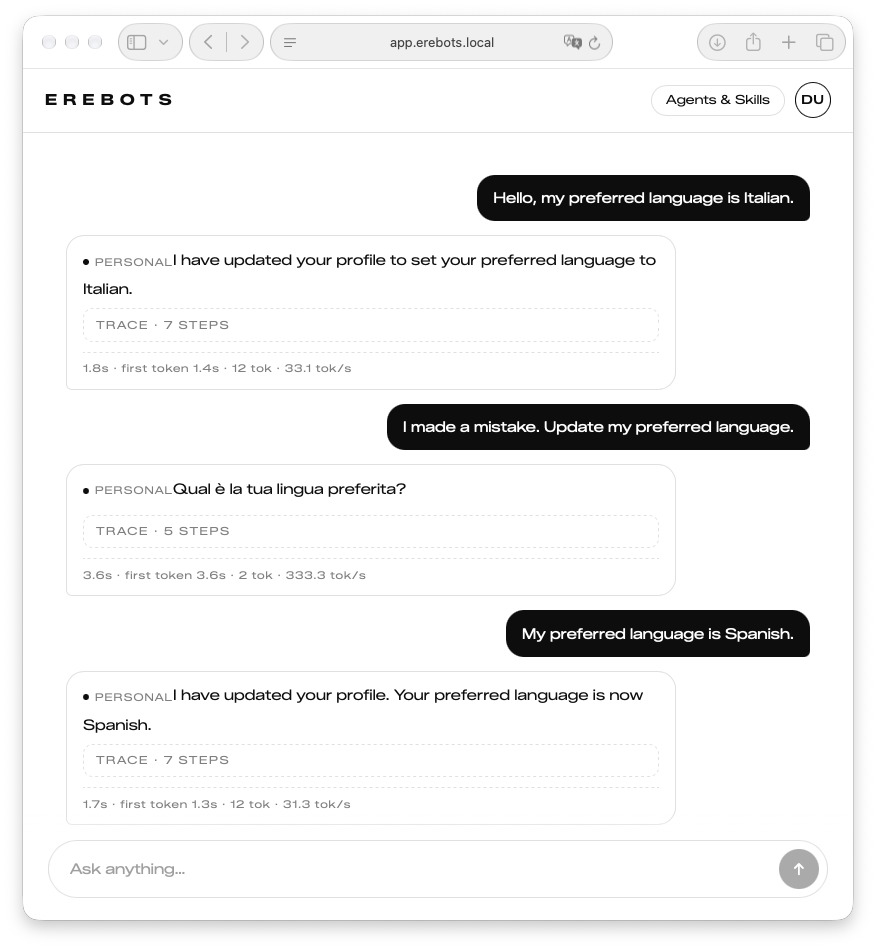}
\caption{The Vue~\cite{vuejs} client during the clarification scenario. The orchestrator routes to \texttt{PersonalAgent}, which asks for the missing value it does not have, and every turn carries its trace and its timings. The question comes back in Italian because the stored preference is already in the prompt.}
\label{fig:web}
\end{figure}

\begin{figure}[H]
\centering
\includegraphics[width=\linewidth]{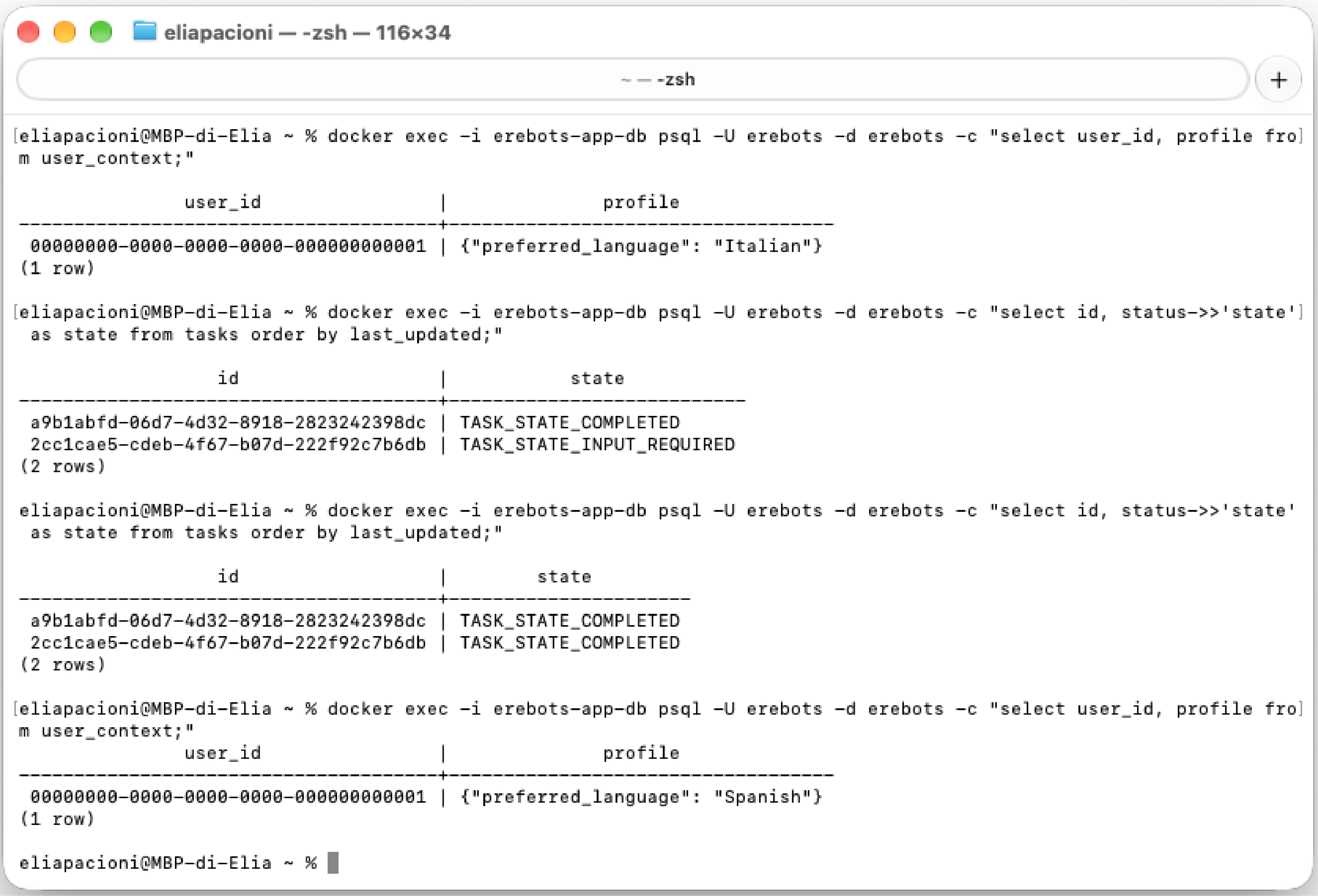}
\caption{The clarification scenario seen in PostgreSQL. The task sits in \texttt{TASK\_STATE\_INPUT\_REQUIRED} while the question is open, completes once the user answers, and the stored profile changes from Italian to Spanish.}
\label{fig:store}
\end{figure}

Figures~\ref{fig:web} and~\ref{fig:store} show one such run on the Docker~\cite{merkel2014docker} stack, with Keycloak, PostgreSQL~\cite{postgresql} and \texttt{a2a\_persistent\_tasks} enabled, driven by \texttt{gemma-4-E4B-it}~\cite{gemma4e4b} over an OpenAI-compatible~\cite{openaiapi} endpoint. The automated suite exercises the same mechanism with stubbed replies, and these figures show it under a model.

\vspace{-15pt}

A user writes \emph{``Remember Italian as my preferred language and convert 12 miles to kilometres.''} The two operations are independent, and routing selects both agents. The conversion returns 19.312128 kilometres from the exact factor 1 mile $=$ 1.609344 kilometres, the trace identifies both A2A calls, and the composed answer keeps the numerical result alongside the outcome of the profile operation. With the conversion endpoint unreachable, the profile operation and its answer survive. The suite covers that fan-out path.

The demonstrator itself starts with authentication disabled and tasks in memory, which leaves the transcript with the client. Reproducing the persisted outcomes needs Keycloak enabled and \texttt{a2a\_persistent\_tasks} set. Measuring the model-dependent half would need a pinned model, fixed prompts, and a decided repetition count, which makes it a study of model behavior.

Routing reaches any endpoint that serves a conforming agent card, so an agent written in another language or on another stack is routable once its card validates. The conformance suite is itself an independent client, built on plain HTTP and JSON Schema with no dependency on \texttt{a2a-sdk}. The direction we have not measured is the reverse, our client against a server built on something other than \texttt{a2a-sdk}.

\section{Impact}
\label{sec:impact}
In EREBOTS 4.0, a single answer from the assistant can entail the work of several agents, local or remote. Adding an agent means listing its URL in a configuration file. That puts questions about routing strategy, delegation policy, and specialist granularity within reach of a researcher unwilling to reimplement discovery, streaming, task lifecycle, and authentication first. Since the framework's mechanics are verified with stubbed model replies, two configurations differ only in the component under study and in the model. Moving an agent to another host takes a different URL, so where a specialist runs becomes a configuration choice.

Conformance pushes EREBOTS beyond custom and possibly unverified agents. The server is measured against the official A2A Technology Compatibility Kit, and the result is reported with the requirements the suite leaves untested. Two research groups that both meet the specification can compose each other's agents with no shared codebase, and the fixtures we ship let either of them run the same check against its own server. 

The framework is self-contained and needs only an OpenAI-compatible endpoint for the model, which a local inference server can serve, and identity and storage run beside it. Given that the composition happens over the wire, a specialist can run where its data and its tools already are, inside the network segment that holds them, while the orchestrator reaches it from outside. The operator therefore decides where the data sits and under which jurisdiction, which is what makes a deployment possible in a regulated domain at all.

Beyond composition, conformance, and self-hosting, the framework has six more advantages: (i) user identity comes from the login and is propagated, with the user's context, to every specialist, while server-side pinning of \texttt{user\_id} prevents a model from reaching another user's data; (ii) exact results come from deterministic MCP tools, not from the model; (iii) a failing specialist removes only its own contribution to the answer; (iv) clarification requests are A2A tasks in the \texttt{input-required} state, resumable only by their owner, before expiry, and across restarts when persisted; (v) traces link every tool call, result, and failure to the delegation that produced it, supporting failure analysis~\cite{cemri2025}; (vi) all frontends share one per-user transcript, which stays consistent under concurrent writes. These properties are enforced by code and do not depend on the model.

However, the framework acknowledges five limitations: (i) it offers only one queued waiting task, so a second clarification in a fan-out waits until it expires; (ii) consent flags are stored but not enforced; (iii) persisted tasks survive restarts of a single instance, but replicas are not coordinated; (iv) specialists do not negotiate; (v) routing and the decision to ask the user depend on the model, and their success rates are not yet measured.

Two lines of work are underway in the group. A nutritional-coaching assistant under the HARRISON~\cite{pacioni2026slr,10.1007/978-3-032-40774-0_36,10.1007/978-3-032-40774-0_17} project continues the earlier EREBOTS work~\cite{calvaresi2021nutrition} as a research prototype, and a second explores industrial use, an assistant for maintaining pharmaceutical manufacturing equipment, with different tools and different specialists.

\section{Conclusions}
\label{sec:conclusions}
In EREBOTS 4.0, the user agent is an endpoint that serves an agent card. The orchestrator uses specialist agents over A2A, routes each request to the matching advertised MCP's kills, and composes their responses. Login, user context, execution traces, and one shared transcript come with the framework. The framework passes 97 requirements of the official A2A conformance suite with no failures.

EREBOTS 4.0 stores consent flags and serves them to the application. No code path refuses an operation when a consent is absent. Enforcing them is harder than in a single-agent assistant, because a consent could cover one operation or a whole specialist, and the check could sit in the orchestrator or in each specialist. For a specialist hosted elsewhere, the decision has to travel with the request, and A2A has no field for it.

Letting one specialist call another is a small change, since each is already an A2A server. Loop detection and a hop limit carried in message metadata would bound the exchanges. The open question is which interaction protocol they should follow. In a contract net~\cite{smith1980contractnet}, agents bid on a task based on their suitability. With LLM-based agents, however, each bid costs a model call, and the reported suitability is itself a model output that may be wrong. Future work will address whether agents' negotiation outperforms orchestrators' routing.

\section*{Acknowledgements} 
This work was supported by the HES-SO RCSO ISNet Hybrid Ai foR RelIable perSonalized cOachiNg (HARRISON) grant, Swiss AI Initiative (a0138).

\bibliographystyle{elsarticle-num}
\bibliography{references}

@article{calvaresi2021erebots,
  author  = {Calvaresi, Davide and Calbimonte, Jean-Paul and Siboni, Enrico and Eggenschwiler, Stefan and Manzo, Gaetano and Hilfiker, Roger and Schumacher, Michael},
  title   = {{EREBOTS}: Privacy-compliant agent-based platform for multi-scenario personalized health-assistant chatbots},
  journal = {Electronics},
  volume  = {10},
  number  = {6},
  pages   = {666},
  year    = {2021},
  doi     = {10.3390/electronics10060666}
}

@inproceedings{calvaresi2021nutrition,
  author    = {Calvaresi, Davide and Eggenschwiler, Stefan and Calbimonte, Jean-Paul and Manzo, Gaetano and Schumacher, Michael},
  title     = {A personalized agent-based chatbot for nutritional coaching},
  booktitle = {IEEE/WIC/ACM International Conference on Web Intelligence and Intelligent Agent Technology (WI-IAT '21)},
  publisher = {ACM},
  year      = {2021},
  pages     = {682--687},
  doi       = {10.1145/3486622.3493992}
}

@article{calvaresi2023slr,
  author  = {Calvaresi, Davide and Eggenschwiler, Stefan and Mualla, Yazan and Schumacher, Michael and Calbimonte, Jean-Paul},
  title   = {Exploring agent-based chatbots: a systematic literature review},
  journal = {Journal of Ambient Intelligence and Humanized Computing},
  volume  = {14},
  number  = {8},
  pages   = {11207--11226},
  year    = {2023},
  doi     = {10.1007/s12652-023-04626-5}
}

@inproceedings{buzcu2024framework,
  author    = {Buzcu, Berk and Pannatier, Yvan and Aydo{\u{g}}an, Reyhan and Schumacher, Michael and Calbimonte, Jean-Paul and Calvaresi, Davide},
  title     = {A framework for explainable multi-purpose virtual assistants: a nutrition-focused case study},
  booktitle = {Explainable and Transparent AI and Multi-Agent Systems (EXTRAAMAS 2024)},
  publisher = {Springer},
  pages     = {58--78},
  year      = {2024},
  doi       = {10.1007/978-3-031-70074-3_4}
}

@misc{a2a2025,
  author       = {{A2A Project}},
  title        = {Agent2Agent ({A2A}) protocol specification, version 1.0},
  howpublished = {\url{https://a2a-protocol.org}},
  note         = {Linux Foundation},
  year         = {2025}
}

@misc{mcp2026,
  author       = {{Model Context Protocol}},
  title        = {Model {Context} {Protocol} Specification},
  howpublished = {\url{https://modelcontextprotocol.io}},
  year         = {2026},
  note         = {Agentic AI Foundation, Linux Foundation}
}

@misc{wu2023autogen,
  author        = {Wu, Qingyun and Bansal, Gagan and Zhang, Jieyu and Wu, Yiran and Li, Beibin and Zhu, Erkang and Jiang, Li and Zhang, Xiaoyun and Zhang, Shaokun and Liu, Jiale and Awadallah, Ahmed H. and White, Ryen W. and Burger, Doug and Wang, Chi},
  title         = {{AutoGen}: Enabling next-gen {LLM} applications via multi-agent conversation},
  year          = {2023},
  url           = {https://arxiv.org/abs/2308.08155}
}

@book{bellifemine2007jade,
  author    = {Bellifemine, Fabio and Caire, Giovanni and Greenwood, Dominic},
  title     = {Developing Multi-Agent Systems with {JADE}},
  publisher = {John Wiley \& Sons},
  year      = {2007}
}

@article{palanca2020spade,
  author  = {Palanca, Javier and Terrasa, Andr{\'e}s and Julian, Vicente and Carrascosa, Carlos},
  title   = {{SPADE} 3: Supporting the new generation of multi-agent systems},
  journal = {IEEE Access},
  volume  = {8},
  pages   = {182537--182549},
  year    = {2020},
  doi     = {10.1109/ACCESS.2020.3027357}
}

@misc{bocklisch2017rasa,
  author       = {Bocklisch, Tom and Faulkner, Joey and Pawlowski, Nick and Nichol, Alan},
  title        = {Rasa: Open source language understanding and dialogue management},
  year         = {2017},
  url          = {https://arxiv.org/abs/1712.05181}
}

@article{pacioni2026slr,
  author  = {Pacioni, Elia and Coman, Andrei C. and Calvaresi, Davide and Manzo, Gaetano and Schumacher, Michael I.},
  title   = {Agent-based hybrid {AI} models and technologies: a systematic literature review},
  journal = {IEEE Access},
  volume  = {14},
  pages   = {21148--21166},
  year    = {2026},
  doi     = {10.1109/ACCESS.2026.3661027}
}

@article{smith1980contractnet,
  author  = {Smith, Reid G.},
  title   = {The contract net protocol: high-level communication and control in a distributed problem solver},
  journal = {IEEE Transactions on Computers},
  volume  = {C-29},
  number  = {12},
  pages   = {1104--1113},
  year    = {1980},
  doi     = {10.1109/TC.1980.1675516}
}

@article{ribeiro2025peak,
  author  = {Ribeiro, Bruno and Dias, David and Gomes, Luis and Vale, Zita},
  title   = {{PEAK}: {Python}-based framework for heterogeneous agent communities},
  journal = {SoftwareX},
  volume  = {30},
  pages   = {102190},
  year    = {2025},
  doi     = {10.1016/j.softx.2025.102190}
}

@misc{botti2025,
  author       = {Botti, Vicent},
  title        = {Agentic {AI} and Multiagentic: are we reinventing the wheel?},
  year         = {2025},
  doi          = {10.48550/arXiv.2506.01463}
}

@article{wooldridge1995,
  author  = {Wooldridge, Michael and Jennings, Nicholas R.},
  title   = {Intelligent agents: theory and practice},
  journal = {The Knowledge Engineering Review},
  volume  = {10}, number = {2}, pages = {115--152}, year = {1995},
  doi     = {10.1017/S0269888900008122}
}

@techreport{fipa2002acl,
  author      = {{Foundation for Intelligent Physical Agents}},
  title       = {{FIPA} {ACL} message structure specification},
  number      = {SC00061G},
  institution = {FIPA},
  year        = {2002},
}

@techreport{rfc6120,
  author      = {Saint-Andre, Peter},
  title       = {Extensible Messaging and Presence Protocol ({XMPP}): core},
  number      = {RFC 6120},
  institution = {Internet Engineering Task Force},
  year        = {2011},
  doi         = {10.17487/RFC6120}
}

@techreport{rfc2119,
  author      = {Bradner, Scott},
  title       = {Key words for use in {RFCs} to indicate requirement levels},
  number      = {RFC 2119},
  institution = {Internet Engineering Task Force},
  year        = {1997},
  doi         = {10.17487/RFC2119}
}

@misc{langgraph,
  author       = {{LangChain, Inc.}},
  title        = {{LangGraph}},
  howpublished = {\url{https://github.com/langchain-ai/langgraph}},
  year         = {2026},
  note         = {MIT licence, accessed 22 October 2026}
}

@misc{crewai,
  author       = {{crewAI, Inc.}},
  title        = {{CrewAI}},
  howpublished = {\url{https://github.com/crewAIInc/crewAI}},
  year         = {2026},
  note         = {MIT licence, accessed 22 October 2026}
}

@misc{cemri2025,
  author       = {Cemri, Mert and Pan, Melissa Z. and Yang, Shuyi and Agrawal, Lakshya A. and Chopra, Bhavya and Tiwari, Rishabh and Keutzer, Kurt and Parameswaran, Aditya and Klein, Dan and Ramchandran, Kannan and Zaharia, Matei and Gonzalez, Joseph E. and Stoica, Ion},
  title        = {Why do multi-agent {LLM} systems fail?},
  year         = {2025},
  url          = {https://arxiv.org/abs/2503.13657}
}

@misc{adk,
  author       = {{Google LLC}},
  title        = {{Agent Development Kit} ({ADK})},
  howpublished = {\url{https://github.com/google/adk-python}},
  year         = {2026},
  note         = {Apache-2.0 licence, accessed 22 October 2026}
}

@misc{microsoft2025agentframework,
  author       = {{Microsoft}},
  title        = {Microsoft Agent Framework: A framework for building, orchestrating and deploying {AI} agents and multi-agent workflows with support for {Python} and {.NET}},
  year         = {2025},
  publisher    = {GitHub},
  journal      = {GitHub repository},
  howpublished = {\url{https://github.com/microsoft/agent-framework}},
  note         = {MIT License, accessed 22 October 2026}
}

@misc{langchain,
  author       = {{LangChain, Inc.}},
  title        = {{LangChain}},
  howpublished = {\url{https://github.com/langchain-ai/langchain}},
  year         = {2026},
  note         = {MIT licence, accessed 22 October 2026}
}

@misc{keycloak,
  author       = {{Keycloak Project}},
  title        = {{Keycloak}},
  howpublished = {\url{https://github.com/keycloak/keycloak}},
  year         = {2026},
  note         = {Accessed 22 October 2026}
}

@article{merkel2014docker,
  title={Docker: lightweight linux containers for consistent development and deployment},
  author={Merkel, Dirk and others},
  journal={Linux j},
  volume={239},
  number={2},
  pages={2},
  year={2014}
}

@misc{vuejs,
  author       = {{Vue.js Core Team}},
  title        = {{Vue.js: The Progressive JavaScript Framework}},
  howpublished = {\url{https://github.com/vuejs/core}},
  year         = {2026},
  note         = {Accessed 26 October 2026}
}

@misc{postgresql,
  author       = {{PostgreSQL Global Development Group}},
  title        = {{PostgreSQL}},
  howpublished = {\url{https://www.postgresql.org}},
  year         = {2026},
  note         = {Accessed 26 October 2026}
}

@misc{gemma4e4b,
  author       = {{Unsloth AI}},
  title        = {{gemma-4-E4B-it}},
  howpublished = {\url{https://huggingface.co/unsloth/gemma-4-E4B-it}},
  year         = {2026},
  note         = {Hugging Face model repository, accessed 26 October 2026}
}

@InProceedings{10.1007/978-3-032-40774-0_36,
author="Pacioni, Elia and Crocetti, Valerio and Coman, Andrei C. and Calvaresi, Davide and Manzo, Gaetano and Schumacher, Michael Ignaz",
title="Autonomous Skill Development: Neuro-Symbolic Agents that Self-generate Verifiable Tools",
booktitle="AIxIA 2026 -- Advances in Artificial Intelligence",
year="2027",
publisher="Springer Nature Switzerland",
address="Cham",
pages="499--513",
isbn="978-3-032-40774-0"
}

@InProceedings{10.1007/978-3-032-40774-0_17,
author="Crocetti, Valerio and Pacioni, Elia and Dragoni, Aldo Franco and Schumacher, Michael Ignaz and Calvaresi, Davide",
title="AiDER: Auditing and Document Evaluation via Rule Compilation and Small Language Models -- An Education Case Study",
booktitle="AIxIA 2026 -- Advances in Artificial Intelligence",
year="2027",
publisher="Springer Nature Switzerland",
address="Cham",
pages="226--240",
isbn="978-3-032-40774-0"
}

@misc{openaiapi,
  author       = {{OpenAI}},
  title        = {{OpenAI} {API} reference},
  howpublished = {\url{https://platform.openai.com/docs/api-reference}},
  year         = {2026},
  note         = {Accessed 26 October 2026}
}

@misc{python313,
  author       = {{Python Software Foundation}},
  title        = {{Python} 3.13 documentation},
  howpublished = {\url{https://docs.python.org/3.13/}},
  year         = {2024},
  note         = {Accessed 26 October 2026}
}

\section*{Current executable software version}
The framework is distributed as source and as a Python wheel, with an optional Docker Compose stack for the full deployment. No standalone executable is distributed, so the optional software metadata table does not apply.

\end{document}